\documentclass{kapproc} 

\usepackage{procps} 

\usepackage[dvips]{graphicx}

\upperandlowercase

\kluwerbib 

\begin{document}

\articletitle{The Chandra X-Ray Observatory}

\articlesubtitle{Observations of Neutron Stars}

\author{Martin C. Weisskopf}
\affil{Space Sciences Department\\
NASA/Marshall Space Flight Center}
\email{martin.c.weisskopf@nasa.gov}

\begin{abstract}
We present here an overview of the status of the Chandra X-ray Observatory which has now been operating for five years. 
The Observatory is running smoothly, and the scientific return continues to be outstanding. 
We provide some details on the molecular contamination of the ACIS filters and its impact on observations. 
We review the observations with Chandra of the pulsar in the Crab Nebula and add some general comments as to the analysis of X-ray spectra. 
We conclude with comments about the future directions for the study of neutron stars with Chandra.
\end{abstract}


\section*{Introduction}
The Chandra X-ray Observatory is the X-ray component of NASA's Great Observatories Program, which today also includes the Hubble Space telescope and the Spitzer Infrared Telescope Facility. NASA's Marshall Space Flight Center (MSFC) manages the project and provides Project Science. The Smithsonian Astrophysical Observatory (SAO) provides technical support and is responsible for ground operations including the Chandra X-ray Center (CXC). 

\subsection{The Observatory}
Work on the Observatory started in 1977 when NASA/MSFC and SAO began the study of what was then named the Advanced X-ray Astrophysics Facility. This study was the outgrowth of NASA's response to an unsolicited proposal submitted in 1976 by Prof. R. Giacconi - Principal Investigator, and Dr. H. Tananbaum - Co-Principal Investigator. In 2002, Prof. Giacconi was awarded the Nobel Prize for his pioneering work in X-ray astronomy. 

\begin{figure}
\begin{center} 
\includegraphics[width=9cm]{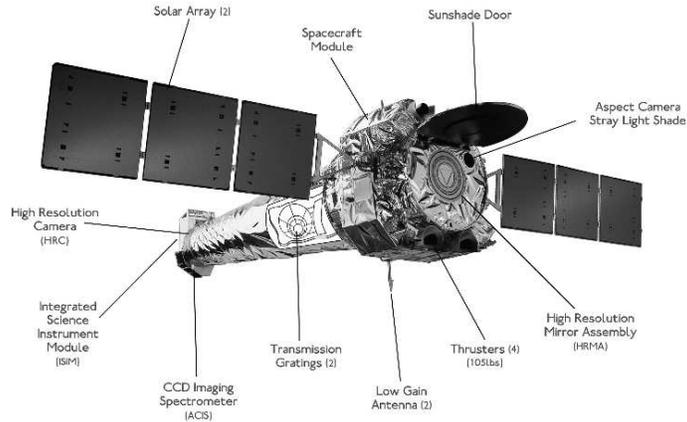}
\caption{Artist's drawing of the Chandra X-ray Observatory.
\label{f:observatory_labeled}}
\end{center}
\end{figure}

The Observatory was launched by the Space Shuttle Columbia on July 23, 1999. The Commander was Col. Eileen Collins, the first female commander of a Shuttle flight. 
With a second rocket system, the Inertial Upper Stage (IUS), attached, the Observatory was both the largest and the heaviest payload ever launched by a Space Shuttle. 
Once deployed, and after separating from the IUS, the flight system, illustrated in Figure~\ref{f:observatory_labeled}, is 13.8-m long by 4.2-m diameter, with a 19.5-m solar-panel wingspan.

The orbit has a nominal apogee of 140,000 km and a nominal perigee of 10,000 km. 
The inclination to the equator is 28.5$^o$ and the satellite is above the radiation belts for more than about 75\% of the approximately 63.5-hour orbital period.

The spacecraft is conventional except for its lightweight construction, which utilizes mostly composite materials. 
The spacecraft provides pointing control, power, command and data management, thermal control, and other such services. 

The specified design life of the mission was 3 years with a goal of 5 which is now being exceeded. 
The only perishable is the gas used for maneuvering and the present supply has the capacity to allow operation for much more than 10 years. 
The orbit will be stable for decades. 

This past year the on-target efficiency was about 65\%. Losses in efficiency are dominated by the time spent in the radiation belts at altitudes below about 60,000 km. 
Other important factors impacting the efficiency are solar activity and maneuver time. 

\subsection{Instrumentation}

The optics and detectors provide sub-arcsecond spectrometric imaging, and, together with two sets of transmission gratings, high-resolution X-ray spectroscopy. The telescope is made of four concentric, precision-figured, superpolished Wolter-1 X-ray telescopes, similar to those used for both the Einstein and Rosat observatories, but of much higher surface quality (more accurate figure and lower surface roughness), larger diameter, and longer focal length which results in better response to higher energies and a larger plate scale.
The mirrors are coated with iridium, chosen for high reflectivity.

The aspect camera system includes a visible-light telescope and CCD camera. A fiducial-light transfer system projects lights attached to the focal-plane instruments onto the aspect camera. 
The aspect solution's accuracy depends on the number of stars detected in the field. 
This number may be as large as 5 and in which case the aspect solution is typically accurate to 0.6 seconds of arc.  

In an assembly off to the side, and just behind the telescope, are 2 objective transmission gratings - the Low-Energy Transmission Grating (LETG), and the High-Energy Transmission Grating (HETG). 
Positioning mechanisms allow one to insert either grating into the optical path to disperse the x-rays onto the focal plane producing high-resolution spectra.
The gratings provide spectral resolving power of E$\delta$E > 500 for wavelengths of > 0.4-nm (energies < 3 keV).

The Space Research Institute of the Netherlands together with the Max-Planck-Institut f\"ur Extraterrestrische Physik designed and fabricated the LETG. The assembly is made of 540 grating facets with gold bars of 991-nm period and provides high-resolution spectroscopy from 0.08 to 2 keV (15 to 0.6 nm).

The Massachusetts Institute of Technology (MIT) designed and fabricated the High-Energy Transmission Grating (HETG) which, in turn, uses 2 types of grating facets - the Medium-Energy Gratings (MEG) which are placed behind the telescope's 2 outermost shells, and the High-Energy Gratings (HEG), behind the 2 innermost shells. 
With polyimide-supported gold bars of 400-nm and 200-nm periods, the HETG provides high-resolution spectroscopy from 0.4 to 4 keV (MEG, 3 to 0.3 nm) and from 0.8 to 8 keV (HEG, 1.5 to 0.15 nm).

The science instrument module includes mechanisms for focusing and translating the two focal-plane instruments: the High Resolution Camera (HRC) and the Advanced CCD Imaging Spectrometer (ACIS). 
SAO designed and fabricated the HRC. 
One of the HRC detectors is a 10-cm-square microchannel plate (HRC-I), and provides imaging over a 31-arcmin-square field of view. 
A second detector (HRC-S), made of 3 rectangular segments (3-cm-by-10-cm each) mounted end-to-end along the grating dispersion direction, serves as the primary readout detector for the LETG. 

The Pennsylvania State University together with MIT built the Advanced CCD Imaging System (ACIS) with charge-coupled devices (CCDs) fabricated by MIT's Lincoln Laboratory. 
As with the HRC, there are two detector systems. 
One is a 2-by-2 array of front-illuminated (FI) CCDs (ACIS-I), and provides high-resolution spectrometric imaging over a 17-arcmin-square field of view. The other (ACIS-S), is a 6-by-1 array mounted along the grating dispersion direction, and serves as the primary readout detector for the HETG. 
Two types of CCDs were used, 4 FI and two back-illuminated (BI) CCDs. The BI CCDs have higher efficiency at lower energies than the FI devices, but were much more difficult to fabricate. 
One BI CCD (ACIS-S3) was placed at the on-axis focal position of the 6 x 1 ACIS-S array. 
Both ACIS detector systems have thin aluminized polyimide filters to minimize contamination by visible light.

\subsection{Complications}

Despite successful science operations, the Observatory has had to deal with some technical difficulties that have had their impact on scientific performance.

\subsubsection{Proton Damage of the FI CCDs}

The front- (not the back-) illuminated CCDs suffered damage which increased the charge transfer inefficiency and thus the energy resolution as a result of bombardment by low energy (100 keV) protons crudely focused by the telescope by means of Rutherford scattering as the Observatory entered the radiation belts. 
A procedure of removing ACIS from the focus position during radiation belt passages has dramatically minimized subsequent degredation of the energy resolution. 
Table~\ref{t:eresolution} gives an indication of the size of the effect for one of the FI CCDs (ACIS-I3) and compares this to the energy resolution for S3 which was unaffected by the low energy protons. 

\begin{table}[ht]
 \caption{Energy Resolution \label{t:eresolution}}
 \begin{tabular}{ccccc}
Energy & Pre-Launch & 2000 & 2004 &   \\
(keV)  & I3         & I3 aim point& I3 aim point& S3 middle\\
\hline
0.5    & 50 eV      & 100 eV      & 104 eV      & 100 eV \\
8.0    & 170 eV     & 370 eV      & 390 eV      & 175 eV\\
\hline
 \end{tabular}
 \end{table}
 
\subsubsection{Molecular Contamination of the ACIS Filters}

\begin{figure}[ht]
\begin{center} 
\includegraphics[width=9cm]{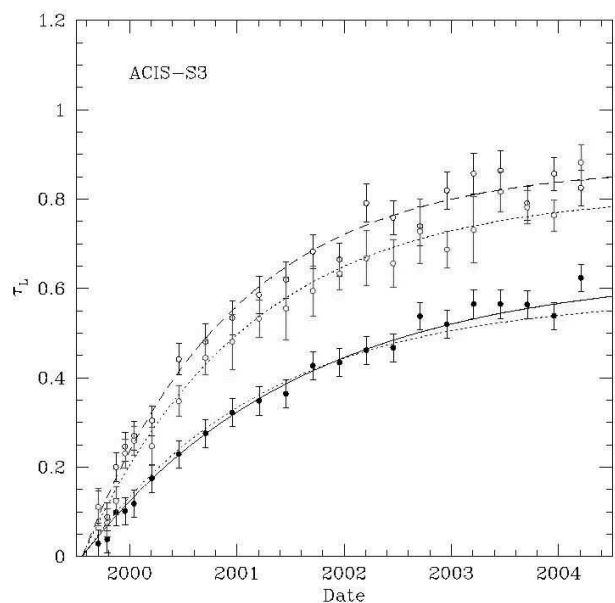}
\caption{Optical depth of the contaminant on the ACIS-S filter versus time based on measurements at the Manganese-L edge. The two upper curves are for the opposite edges of the filter in the long direction. The lower curve is for the center of the filter. The increased optical depth at the edges reflects the temperature gradient. Figure kindly provided by A. Vikhlinin (CXC).
\label{f:contamination}}
\end{center}
\end{figure}

Both ACIS filters, which are close to the CCDs and therefore near the coldest (120 C) surfaces on the observatory, are collecting hydrocarbon contamination at the rate of about one-half an optical depth at the carbon K-edge per year. 
Figure~\ref{f:contamination} shows the buildup as a change in the optical depth to the Manganese-L complex of the ACIS calibration source at about 0.7 keV.  
Using astronomical sources and grating spectra one finds C, O, and F edges. 
The composition is C:O = 11.5:1 and C:F is about 14:1. 

Prior to launch it had been planned to bake off the expected contamination after a nominal amount of buildup. 
Post launch, however, the prospect of bake out was complicated by its potential negative impact on the charge transfer efficiency of the FI CCDs. 
These appeared to suffer an increase in charge transfer inefficiency (and thus an increase in energy resolution) as a consequence of an early bakeout. 
The Chandra Project is performing detailed studies to determine if there is an effective bakeout strategy.  
Buildup of contamination on the ACIS filters has had various impacts on the science program. Observations of any object for which the absorbing column is greater that about $5 \times 10^{21}$ cm$-2$ are not impacted. Conversely, the soft part of the spectrum of unabsorbed sources has become more and more difficult to observe.

Many more details on the instrumentation and the performance on-orbit may be found in the paper Weisskopf et al. (2004a). 

\section*{Observations of the Pulsar in the Crab Nebula}

\begin{figure}[ht]
\begin{center} 
\includegraphics[width=9cm]{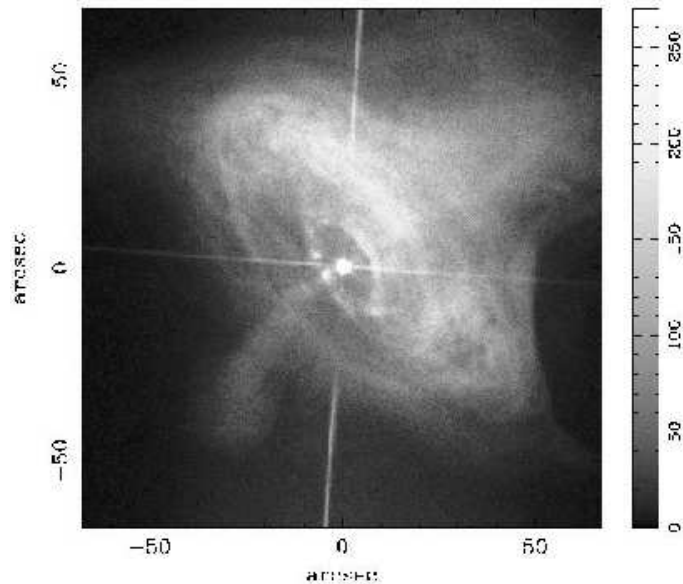}
\caption{LETG/HRC-S zero-order image of the Crab Nebula. The two streaks are due to dispersion of the image of the pulsar (including cross-dispersion) by the gratings. 
\label{f:crab}}
\end{center}
\end{figure}

Perhaps for more than any other object, the Chandra observations of the Crab Nebula and its pulsar illustrate the power of the Observatory for furthering our understanding of neutron stars. 
Figure~\ref{f:crab} shows a version of the now famous image (Weisskopf et al. 2000) which revealed so dramatically the extraordinarily complex spatial structure in the emission from the nebula. 
The bright inner ring is direct evidence for the shock front that forms where the wind of particles from the pulsar start to radiate via the synchrotron process. 
For more details on the pulsar wind nebula see Hester et al. (2002).

Exploiting the angular resolution of the Observatory was also crucial to avoid the large background produced by the nebula and thus enabling the discovery that the pulsar was always on, even at the so-called pulse minimum --- Figure~\ref{f:crab_minimum} --- see Tennant et al. 2001 for details.

\begin{figure}[ht]
\begin{center} 
\includegraphics[width=9cm]{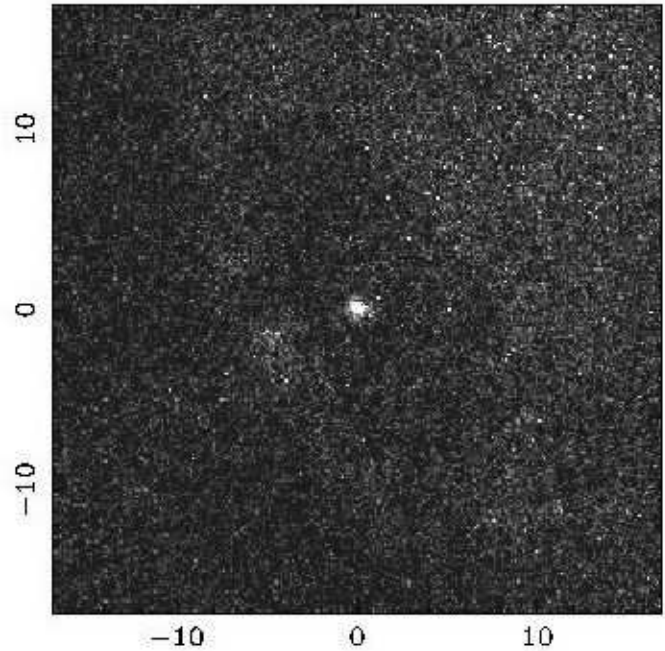}
\caption{LETG/HRC-S zero-order image of the Crab Nebula at pulse minimum. The horizontal and vertical axis are in arc-seconds with the origin at the pulsar. 
\label{f:crab_minimum}}
\end{center}
\end{figure}

Being able to resolve the pulsar from the nebula also allowed one to study the spectrum as a function of pulse phase (Weisskopf et al. 2004b).
Interestingly, the best fit spectrum for the pulsar (0.3 to 4.2 keV) when integrated over all pulse phases is, indeed, a power law with spectral parameters: $\Gamma_{\rm P}= 1.643\pm0.020$, $N_{\rm H} = (3.77\pm0.12) \times 10^{21}$, and [O/H] = $(3.18\pm0.30) \times 10^{-4}$.
Weisskopf et al. (2004b) also discuss the impact on the spectral parameters of utilizing various different cross sections and abundances. 
In performing these fits we allowed for the effects of scattering by the interstellar medium, a consideration that one must pay attention to when using Chandra.
We found, amongst other things, that the spectral models using Angers and Grevesse (1989) never fit the data unless oxygen was underabundent. 
Models using Wilms, Allen, and McCray (2000) fit the data with oxygen only slightly underabundent. 
(Not a contradiction as the best-fit abundance was the same within the errors.) 
The reader should refer to the paper by Mori et al. (2004) for an excellent discussion of the spectroscopy of the nebula.

\begin{figure}[ht]
\begin{center} 
\includegraphics[width=9cm]{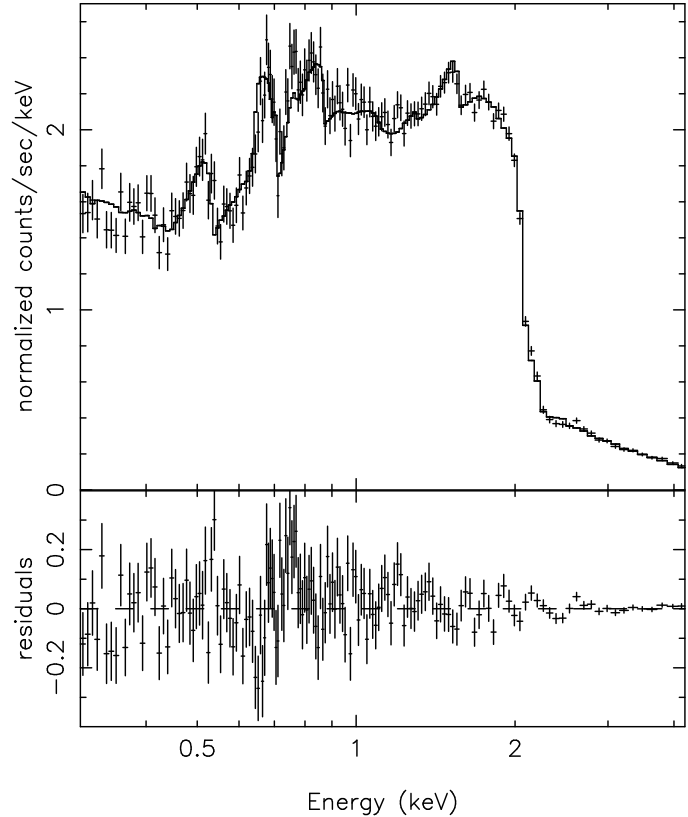}
\caption{Chandra LETGS spectrum of the Crab pulsar compared with a model using the abundances of Anders and Grevasse (1989) and the cross-sections of Balucinska-Chrurch and McCammon (1992) with He cross section from Yan, Sadeghpour, and Dalgarno (1998) and with the oxygen column set by the data. 
\label{f:crab_minimum}}
\end{center}
\end{figure}

\begin{figure}[ht]
\begin{center} 
\includegraphics[width=8cm, angle=-90]{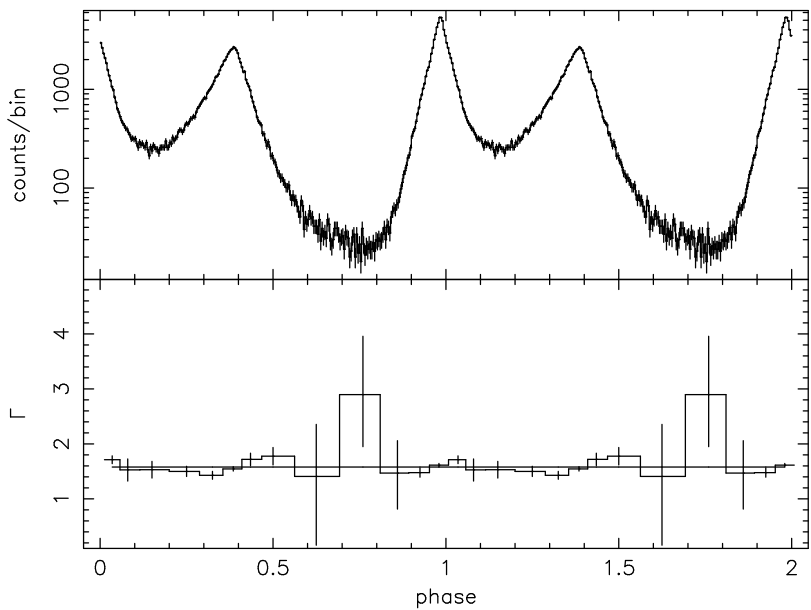}
\caption{The light curve (upper) and photon index (lower) as a function of pulse phase for the Crab pulsar. Two cycles are plotted for clarity. 
\label{f:crab_spectrum_phase}}
\end{center}
\end{figure}

\begin{figure}[h]
\begin{center} 
\includegraphics[width=8cm, angle=-90]{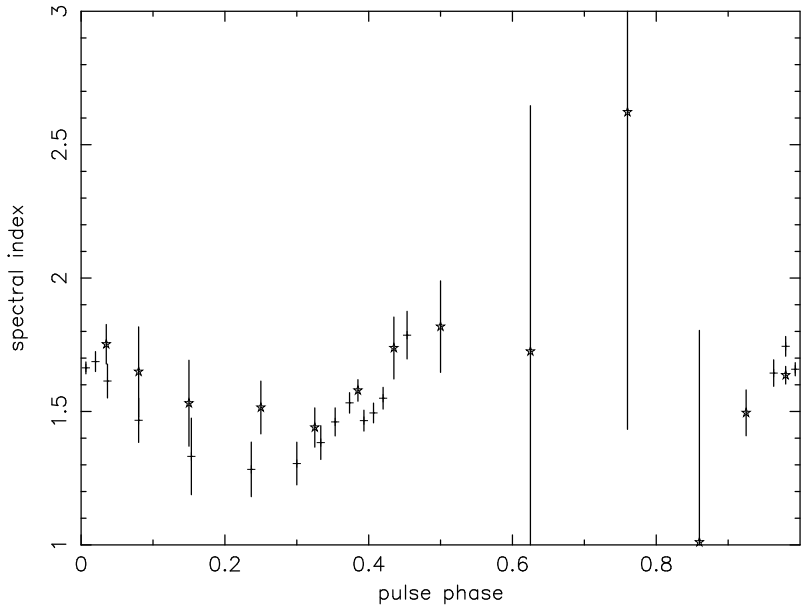}
\caption{Power law spectral index as a function of pulse phase comparing the Chandra measurements (the stars) with results obtained using Beppo-Sax. 
\label{f:comparison}}
\end{center}
\end{figure}

Figure ~\ref{f:crab_spectrum_phase} shows the first measurements of the spectrum of the pulsar at {\em all} pulse phases. 
These data were used to reject, but only at 85\%-confidence, the hypothesis that the spectral index is constant.
Note that the apparent variation of spectral index between phases -0.1 and 0.5 is qualitatively similar with other measurements, e.g. as in Figure~\ref{f:comparison} (Massaro et al. 2000, Pravdo, Angelini, and Harding 1997) 

Using assumed properties (equation of state, mass,
radii, etc.) for a 1.358-$M_\odot$ neutron star with $R_{\infty}$ = 15.6~km~--- appropriate for models that assume neutron-star cores with moderately stiff equations of state and containing strong proton superfluidity~---  we set limits to the surface temperature. 
For a distance of 2~kpc, we found a 3-$\sigma$ upper limit to the (gravitationally-redshifted) blackbody surface temperature viewed at infinity of $T_{\infty} <  1.97$~MK [$\log T_{\infty} < 6.30$]. 
We regard $T_{\infty}$ as a representative and indicative upper limit.

\section*{Spectral Fitting}
In performing the spectral fits to the Crab pulsar, we were reminded of several points that are important when trying to determine spectral parameters.
A review of the current literature reinforces our opinion that these points need to be re-emphasized from time to time. 
The first point is that authors should inform readers which cross sections and abundances were used. 
We recommend that the cross-sections discussed by Wilms, Allen and McCray be considered as they are modern and allow for absorption (but not diffractive scattering) due to the interstellar dust grains. 
Ultimately, diffractive scattering will need to be added.
 
A second point is that one should report the quality of the fit of the model(s) to the data and discuss how  uncertainties were determined. 
It is our impression that it may not be understood by all that, if the fit is poor, the uncertainties automatically spewed forth by modern fitting engines are {\em underestimates}. 
Redefining the uncertainties in the data points to yield a value for the chi-squared statistic of unity, and then determining extremes on confidence contours is an appropriate approach to estimating uncertainties in these circumstances.

\begin{figure}[ht]
\begin{center} 
\includegraphics[width=8cm, angle=-90]{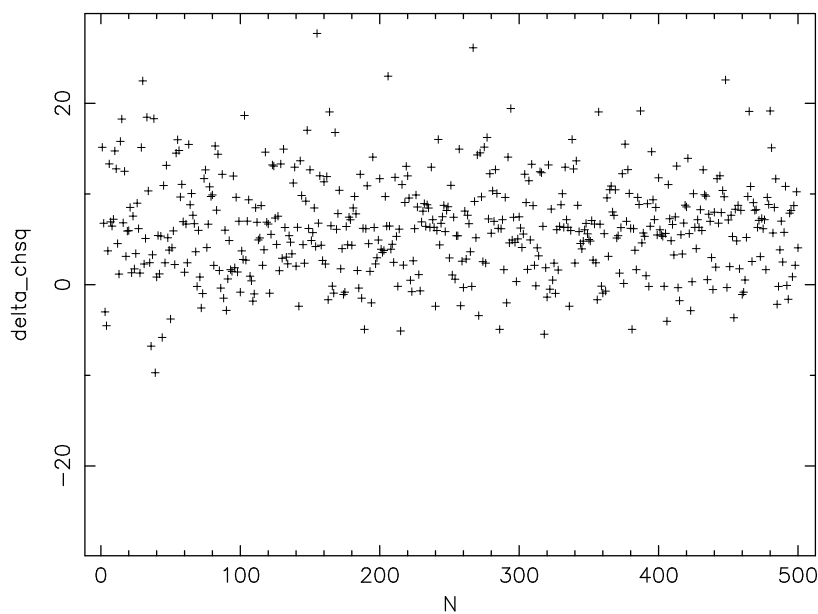}
\caption{The difference in the chi-squared statistic for 500 trials of fitting powerlaw-simulated data to a powerlaw and to a blackbody. The powerlaw model consistently results in the lower value of chi-squared despite being statistically indistinguishable from the blackbody. 
\label{f:deltachi}}
\end{center}
\end{figure}

It is also interesting to ask why is it, with the advent of the CCD spectra available from Chandra and XMM-Newton, that all too often one does not seem to be able to distinguish between different physical models for the spectrum?
Part of the answer to this question is that chi-squared is a zeroth order statistic. 
By way of demonstrating this point we used Monte-Carlo simulations to simulate Chandra data where the true underlying spectrum was a powerlaw and then fit these data to both powerlaw and blackbody models. For the simulations in question the two values of chi-squared (one for each spectral model) were statistically indistinguishable and yet the value obtained from fitting the powerlaw was almost always lower than that obtained from fitting to the blackbody spectrum.  
Figure~\ref{f:deltachi} shows that only 63 of 500 measurements had a difference in chi-squared of less than zero.

We don't (yet) have an answer to the question we posed, but it seems clear that one should define a statistic that is more sensitive to the astrophysical question of interest and then tailor the observing strategy to optimize the sensitivity to that statistic. 

\section*{Future Observations of Neutron Stars with Chandra}
We should take advantage of high-resolution imaging in the X-ray band while we have it. 
I would expect to see more concentrated efforts on providing deep, spectrally-resolved, images of pulsar wind nebulae in order to further probe the interaction of the neutron star with its surroundings.
I would anticipate that we shall learn much by studying the deep probes of the galactic center and comparing statistical properties of luminosities, etc. with similar results obtained from deep exposures of nearby galaxies.
These data will be invaluable for decades to come.

\begin{acknowledgments}
The author wishes to thank all the scientists involved with the Chandra project for their contributions to the success of this Great Observatory. 
\end{acknowledgments}

\end{document}